\documentclass[referee]{aa} 
\usepackage{epsfig}
\usepackage{natbib}
\usepackage{graphicx}
\usepackage{txfonts}

\newcommand{\bea}{\begin{eqnarray}}    
\newcommand{\eea}{\end{eqnarray}}      
\newcommand{\be}{\begin{equation}}
\newcommand{\ee}{\end{equation}}
\newcommand{\bef}{\begin{figue}}
\newcommand{\eef}{\end{figure}}

\def\spose#1{\hbox to 0pt{#1\hss}}
\def\ltapprox{\mathrel{\spose{\lower 3pt\hbox{$\mathchar"218$}}
\raise 2.0pt\hbox{$\mathchar"10C$}}}
\def\gtapprox{\mathrel{\spose{\lower 3pt\hbox{$\mathchar"218$}}
\raise 2.0pt\hbox{$\mathchar"10E$}}}
\def\inapprox{\mathrel{\spose{\lower 3pt\hbox{$\mathchar"218$}}
\raise 2.0pt\hbox{$\mathchar"232$}}}

\begin{document}
   \title{Gravitational fluctuations of the galaxy distribution}

   \subtitle{}

   \author{Francesco Sylos Labini} \titlerunning{Gravitational
     fluctuations of the galaxy distribution} \authorrunning{Sylos
     Labini} \institute{Enrico Fermi Center, Piazza del Viminale 1,
     00184 Rome, Italy, and Istituto dei Sistemi Complessi CNR, Via
     dei Taurini 19, 00185 Rome, Italy}

   \date{Received xxxx; accepted xxxx}

 
  \abstract {} {We study the statistical properties of the
    gravitational field generated by galaxy distribution observed by
    the Sloan Digital Sky Survey (DR7). We characterize the
    probability density function of gravitational force fluctuations
    and relate its limiting behaviors to the correlation properties of
    the underlying density field.  In addition, we study whether the
    PDF converges to an asymptotic shape within sample volumes.}  {We
    consider several volume-limited samples of the Sloan Digital Sky
    Survey and we compute the gravitational force probability density
    function (PDF).  The gravitational force is computed in spheres of
    varying radius as is its PDF. } { We find that (i) the PDF of the
    force displays features that can be understood in terms of galaxy
    two-point correlations and (ii) density fluctuations on the
    largest scales probed, i.e. $r\approx 100$ Mpc/h, still contribute
    significantly to the amplitude of the gravitational force.}  {{Our
      main conclusion is that fluctuations in the gravitational force
      field generated by galaxy structures are also relevant on scales
      $\sim 100$ Mpc/h.  By assuming that the gravitational
      fluctuations in the galaxy distribution reflect those in the
      whole matter distribution, and that peculiar velocities and
      accelerations are simply correlated, we may conclude} that
    large-scale fluctuations in the galaxy density field may be the
    source of the large-scale flows recently observed.}


\maketitle
%

\section{Introduction}

{ One of the main problems in studying the dynamics of a system of
  point masses concerns the analysis of the force acting on a test
  particle. This analysis can be achieved by determining of the
  statistical properties of the (Newtonian) gravitational field
  generated by the given point-mass distribution. From this study in
  general one can derive several useful pieces of information about
  the statistical and dynamical properties of the given point-mass
  distribution
  \citep{chandra43,atrio,force,masucci,scaleforce,lattice,book}.  It
  has been known since the pioneering article of \citet{chandra43}
  that, when particles are distributed without correlations (i.e., a
  Poisson distribution), the second moment of the gravitational force
  diverges.  Certain spatial correlations between particles may cause
  the divergence, for instance, of the first moment \citep{force}. For
  this reason, instead of computing the moments of the gravitational
  force, it is more appropriate to determine its probability density
  function (PDF). In particular, to understand the relevant features
  of the gravitational force fluctuations, it is interesting to study
  the limiting behaviors of the PDF in both strong and weak fields. It
  can be shown (see below) that the former is closely related to the
  small scale two-point correlation properties of the distribution. On
  the other hand, the weak field behavior is determined, in a
  non-trivial way, by how rapidly the distribution becomes isotropic
  on large scales. Therefore, this latter behavior is connected to the
  higher order correlation properties of the distribution
  \citep{force,book,lattice}.

Our first aim, in this paper, is to consider the determination of the
PDF of the gravitational force generated by the observed spatial
distribution of galaxies. Given that the galaxy distribution is
described as a stochastic point process, the statistical properties of
the gravitational field on an arbitrary point is itself a stochastic
quantity. The properties of these two stochastic processes are closely
related.  In particular, the force PDF can be simply regarded as a
particular statistical characterization of the galaxy distribution
containing some useful statistical information about higher order
galaxy correlations.

To connect the force PDF of the galaxies with that of the whole mass
distribution, we have to make an hypothesis about the relation between
luminous and dark matter. The most simple assumption is that the
galaxies trace the peaks of the underlying matter density field, i.e.,
that the galaxy density field is linearly biased with respect to the
dark matter density field. If this is so, then the fluctuations in the
gravitational force in the galaxy distribution also reflect those in
the full matter distribution.

This is actually the standard hypothesis that is usually assumed in
studies that compare of the dipole induced by the gravitational
influence of structures in our local Universe with the observed CMB
dipole~\footnote{ { This shows that our Local Group moves with a
    velocity of 627 $\pm$ 22 km/ s towards the direction $l$ =
    276$^{\circ}$ and $b$ = 306$^{\circ}$ in Galactic coordinates.}}.
Nowadays, several groups have shown that structures responsible for
the dipole must be located on scales larger than $\sim 150$ Mpc/h
\citep{maller}.  However, because of the sparseness of data at very
large distances and the incompleteness of galaxy surveys, there is
still no consensus about the depth of the convergence for the CMB
dipole \citep{lavaux2008}.

The actual dipole caused by structures around us corresponds to a
determination of the force around a single point of the galaxy
distribution (our galaxy).  On the other hand, the PDF is determined
by computing the gravitational force on each galaxy contained in a
given catalog, and by repeating this measurement for all measured
galaxies.  With respect to the single determination, the PDF has
several advantages: (i) it considers complete samples extracted from
galaxy redshift surveys (i.e., volume-limited samples) without making
any additional assumption about the incompleteness of the
catalogs. (ii) We do not have to consider the sky regions where
observations were not done and use weighting schemes to correct for
this incompleteness. We indeed compute the force only in spherical
volumes fully contained in our samples. (iii) By considering each
galaxy as a center, we determine the force generated by all galaxies
contained in a spherical volume around it. We can therefore evaluate
the statistical significance that the determination of the force from
a single point (i.e., the dipole) is (or is not) found to converge to
an asymptotic value within a certain distance.

It is then interesting to consider the relation between the
gravitational force field $\vec{g}$ and the peculiar velocity field
$\vec{v}$.  In the linear regime of low-amplitude density
fluctuations, it is well-known that the velocity is proportional to
the gravitational acceleration \citep{pee80}.  As our observations are
limited to a range of scales in which galaxy fluctuations are still
large, we can only speculate about the relation between the
gravitational force and the peculiar velocity fields by {\it assuming}
that there is a direct proportionality even in the non-linear
regime. In this respect we note that when fluctuations are large and
the universe is curvature-dominated (i.e., with negligible matter
content) one still finds that $\vec{v} \propto \vec{g}$
\citep{pwa}. The relation between $\vec{g}$ and $\vec{v}$ has acquired
interest because of the increasing number of observations of peculiar
velocities on large scales.

Peculiar velocities provide important dynamical information because
they are related to the large-scale matter distribution.  By studying
their local amplitudes and directions, these velocities allow us, in
principle, to probe deeper, or hidden, parts of the Universe.  The
peculiar velocities are indeed directly sensitive to the total matter
content, inferred from its gravitational effects, and not only to the
luminous matter distribution. However, their direct observation by
means if distance measurements remains a difficult task.  There have
been observations of large-scale galaxy coherent motions that are at
odds with standard cosmological models. For instance,
\cite{watkins2008} measured in a Gaussian window of diameter 100
Mpc/h, a coherent bulk motion of $407 \pm 81$ km/s, which conflicts
with the prediction of $\approx 200$ km/s at the 2$\sigma$ level of
LCDM.  Other independent results confirm the presence of unexpectedly
large bulk motions \citep{lavaux2008,kaslinski2008,kaslinski2009}.  In
particular, \citet{kaslinski2008} measured peculiar velocities of
galaxy clusters by studying, in a large statistical sample of
clusters, the fluctuations in the cosmic microwave background (CMB)
generated by the kinematic Sunyaev-Zeldovich (KSZ) effect, i.e.,
caused by the scattering of microwave photons by the hot X-ray
emitting gas inside clusters. They measured a bulk flow of 600-1000
km/s on scales $>$ 300 Mpc/h (the limit of the considered clusters
catalog) which is more than 10 times larger than expected by the
LCDM model.  In addition, they concluded that the coherence length of
the measured bulk flow shows no signs of convergence out to scales.
\citet{kaslinski2009}, by applying the KSZ method to a larger cluster
sample, showed that the flow is consistent with approximately constant
velocity out to at least $\simeq 800$ Mpc/h. Although these results
remain controversial \citep{keisler}, measurements of peculiar
velocities have very interesting potentials and future surveys are
expected to provide more precise data of bulk flow motion on large
scales \citep{smith04,frieman08}.  Our aim here is to verify whether,
based on the assumption that $\vec{v} \propto \vec{g}$, fluctuations
in the force distribution should be compatible with the large-scale
bulk flows observed. If this were so, this would indicate that local
galaxy structures may represent an important contribution to the
peculiar velocities that could avoid us having to invoke exotic
physics, as the effect of far-away pre-inflationary inhomogeneities,
to explain the large-scale bulk flows \citep{kaslinski2008}.

In this paper, we focus on the following two questions: (i) What is
the shape of the PDF and how is this connected to the underlying
fluctuations in the galaxy density field ? (ii) Is there a convergence
of the PDF to an asymptotic shape that is independent of the size of
the volume (i.e., the sphere radius $r$) in which this is
computed. Based on the assumption that light traces mass, we are
unable to determine the direction and amplitude of the dipole around
us, thus verify whether this is the same as the CMB dipole. However,
we can characterize the properties of gravitational fluctuations
generated by galaxy structures in a complete statistical way (up to
the maximum scale $R_s$ allowed by the geometry of available samples);
in particular, we can determine whether in the volumes considered the
gravitational force has reached, from a statistical point of view,
saturation on scales $r_f <R_s$. In this case, fluctuations on scales
larger than $r_f$ do not contribute anymore, statistically, to the
amplitude of the gravitational force. That is, large-scale
contributions to the force from different directions tend to
compensate each other.
The scale $r_f$ would thus be the scale beyond which large scale flows
would start to decay in amplitude. On the other hand, if we were able
to place only a lower limit on the scale $r_f$, i.e. if $r_f > R_s$
without a convergence of PDF to a stable shape within the sample
volume, this would corroborate the conjecture that large-scale flows
can be sourced by fluctuations in the density field that statistically
break spherical symmetry. 
In this case, larger and larger volumes would continue to contribute
to the gravitational force on an average galaxy. Therefore, if the
measurements of large-scale bulk flows are unaffected by major
statistical and systematical effects, but are generated instead by
galaxy density inhomogeneities, we would expect not to see a
convergence of the PDF to a stable shape inside the considered samples
because we are limited to a volume of radius $R_s\approx 100$ Mpc/h.
As discussed above, while the information on the force PDF of galaxies
can be derived in straightforward way from the data, its relation to
the peculiar velocity field is based on a series of
assumptions. Nevertheless, it is interesting to consider whether when
galaxies trace mass and peculiar velocities are parallel to
accelerations, it is possible to learn something about the relation
between the force and peculiar velocity fields.

The paper is organized as follows. In Sect.\ref{gfspd}, we briefly
review some basic properties of the gravitational force field in
stochastic point distributions. We then describe the selection of the
galaxy samples in Sect.\ref{data} and the methods (Sect.\ref{methods})
used to estimate the relevant statistical quantities in the data.  In
Sect.\ref{results}, we discuss our main results and in
Sect.\ref{mocks} we consider the PDF of the gravitational force in
mock galaxy catalogs generated from cosmological N-body
simulations. We finally draw our main conclusions in
Sect.\ref{conclusions}.  }

\section{Gravitational force from stochastic point distributions} 
\label{gfspd} 

The value of the gravitational force acting on a particle belonging to
a given point distribution is determined by both the immediate
neighborhood of the particle itself and the large-scale properties of
the system. This situation reflects two peculiar features of the
gravitational force, that it is divergent at small separations and is
slowly, i.e., in accordance with a power-law, decaying at large
distances. The stochastic nature of a given point distribution
generates a certain PDF $P(\vec{F})$ of the gravitational force
$\vec{F}$. In general, one would like to calculate $P(\vec{F})$ from
the statistical properties of the underlying particle
distribution. There are a few examples of particle distributions with
analytically calculable $P(\vec{F})$.  \citet{chandra43} was the first
to compute the PDF of the gravitational force for an uncorrelated
distribution of points, i.e. the Poisson distribution. In this
situation, the PDF of the force is found to be given exactly by the
Holtzmark distribution (HD), which is a three-dimensional fat (or
heavy) tailed L\'evy distribution.  Approximated generalizations of
this approach can be found for more complex particle systems obtained
by perturbing a Poisson distribution \citep{masucci} or perturbing a
perfect cubic lattice of particles \citep{lattice}.  \citet{vlad}
developed a functional integral approach for evaluating the stochastic
properties of vectorial additive random fields generated by a variable
number of point sources obeying inhomogeneous Poisson statistics.  In
this way he derived the generalized Holtzmark distribution (GHD),
which we discuss in more detail below.

The  gravitational force  for  unit  mass due  to the $N$   particles 
contained in a  sphere $C(r,\vec{x}_i)$ of radius $r$  centered on the
$i^{th}$ particle of a given distribution is
\be 
\label{force}
\vec{F}_i (r) =  \sum_{j=1; j \ne i; j\in C(r,\vec{x}_i)}^{N} G m_j
\frac{\vec{x}_i-\vec{x}_j}{|\vec{x}_i-\vec{x}_j|^3} 
\ee 
where $G$ is the gravitational constant, $m_j$ is the $j^{th}$
particle mass, and $N$ is the number of particles in a given volume
$V$. In what follows, we take $G=1$ and we consider particles with the
same mass, so that we assume that $m_i=1 \;\; \forall \;\; i$.

If the particle distribution is statistically stationary, i.e.  it is
invariant under space rotations and translations, the direction of
$\vec{F}$ has equal probability in each direction, i.e.  $P(\vec{F})$
depends only on the force modulus $F=|\vec{F}|$. For this reason, we
can consider the PDF $W(F)$ of the force modulus $F$, which is simply
related to the three dimensional PDF by
%
$W(F) = 4 \pi F^2 P(\vec{F}) \;. $
%

The statistical properties of a density field can be characterized
by measuring the average conditional density defined as
\citep{sdss_aea} 
\be
\label{apd2}
\langle n(r) \rangle_p = \frac{1}{N} \sum_{i=1}^{N} n_i(r) = A r^{D-3}
\;, \ee
where $n_i(r)$ is the density measured in a sphere of radius $r$
centered on the $i^{th}$ particle, and the second equality holds only
for the case in which there are power-law correlations in a certain
range of scales. In Eq.\ref{apd2}, the constant $A$ is related to the
average nearest neighbor (NN) distance $\Lambda \approx A^{-1/D}$ and
$\gamma=3-D$ is the correlation exponent \footnote{Note that $D$ is
  not the space dimension, but it can be interpreted as the fractal
  dimension of the object \citep{book}.} . When there are power-law
correlations on small scales one can readily derive an expression for
the force PDF due only to a NN particle by using the identity
~\citep{book}
\be  
\label{wnnp}
W_{nn}(F) |d F| = \omega(r) |dr| \;, 
\ee  
where 
\be
\label{nnapd}
\omega(r) = 4 \pi A r^{D-1} \exp\left(-\frac{4 \pi A}{D} r^{D} \right)
\ee
 is the NN PDF for a particle system obeying to Eq.\ref{apd2}
\footnote{ { More precisely, Eq.\ref{nnapd} is found when higher order
    correlations are neglected.}}.  For $D=3$, Eq.\ref{nnapd} reduces
to the well-known expression for the Poisson NN PDF \citep{chandra43}.
Thus, from Eq.\ref{wnnp}-\ref{nnapd} we can derive the NN force PDF
\be
\label{eqWnnD1}
W_{nn}(F;D) = \frac{D}{2} F_\Lambda^{D/2} F^{-\frac{D+2}{2}} 
\exp\left[ - \left(\frac{F_\Lambda}{F}\right)^{D/2} \right] \;, 
\ee
where 
$F_\Lambda =  \left( \frac{4 \pi}{D} A \right)^{2/D} \;. $ 
%
For $D=3$, Eq.\ref{eqWnnD1} reduces to the Poisson case
\citep{chandra43}.  The asymptotic behavior in the strong field limit
is
\be
\label{eqWnnD3}
\lim_{F \rightarrow \infty} W_{nn}(F;D) \rightarrow \frac{D}{2}
F_\Lambda^{D/2} F^{-\frac{D+2}{2}} \;.  \ee
It is simple to show that the average force diverges for $D>2$ and
that the average square force diverges $\forall D\le 3$.  { For this
  reason, as mentioned in the introduction,} we do not directly
consider the study of these moments, while we focus on the shape of
the PDF.

\citet{vlad} derived a simple expression for the GHD, which holds when
there are only power-law two-point correlations as in Eq.\ref{apd2}
and no higher order ones (see also the discussion in
\citet{force}). The GHD can be written as
\be
\label{eqHD1}
W(F;D) = \frac{H(\beta;D)}{F_0} \;, 
\ee
where $\beta=F/F_0$, 
\be
\label{eqHD2}
H(\beta;D) = \frac{2}{\pi\beta} \int_0^{\infty} 
\exp\left(-(x/\beta)^{D/2} 
\right) x \sin(x) dx \;, 
\ee
the constant $F_0$ is given by 
\be
\label{eqHD3}
F_0 = \left( \frac{\pi} { (D+2) \Gamma_E \left(\frac{D}{2}\right)
  \sin\left(\frac{D\pi}{4} \right)} \right)^{2/D} F_\Lambda \;, 
\ee
and the symbol $\Gamma_E$ represents the Euler Gamma function.  For
$D=3$, Eqs.\ref{eqHD1}-\ref{eqHD3} reduces to the HD.  It is easy to
show that in the strong field limit $ W(F;D) \rightarrow W_{nn}(F;D)$,
so that the fat tail is always present and due to NN correlations.

The PDF of the vector $\vec{F}$, i.e.  $P(\vec{F})$, must be
symmetrical about zero, because the distribution is isotropic and the
average of the vector force must be zero. Thus $P(\vec{0}) =const.$,
so that we have
\be
\label{eq:wf1}
\lim_{F\rightarrow 0} W(F) \propto F^2P(0) \propto F^2 \;,
\ee
which is indeed the case for both the HD and the GHD. Hence the
limiting behaviors of the GHD (and of the HD for $D=3$) are 
\bea 
\label{limits}
&& W(F) \propto F^{-(D+2)/2} \;\;\mbox{for}\;\; F\rightarrow \infty \;, 
\\ \nonumber 
&& W(F) \propto F^{2} \;\;\mbox{for}\;\; F\rightarrow 0
\;.
\eea 
The strong field limit is determined by the small-scale correlation
properties, while the weak field limit is very general and related to
the stationarity of the distribution.  We focus on these tails in the
analysis of the galaxy samples.

We emphasize that Eq.\ref{apd2} may describe an highly inhomogeneous
structure characterized by large fluctuations and non-trivial higher
order correlations \citep{book}.  In this a situation neglecting
higher order correlations is a very crude approximation. However, only
in this limit is one able to calculate the force PDF. By taking into
account higher order correlations, one may compute only few moments of
the force \citep{force,book}.  This means that, while the limiting
behaviors given by Eq.\ref{limits} are expected to be satisfied as
long as Eq.\ref{apd2} holds and the particle distribution is
stationary, the exact location of the peak, marking the transition
from $W(F) \propto F^{-(D+2)/2}$ to $W(F) \propto F^2$, is poorly
constrained from a theoretical point of view.

{ We note that Eq.\ref{apd2} may also describe a structure that has
  varying correlations as a function of scale. In this case, $D=D(r)
  \le 3 $ where $D=3$ when the distribution becomes spatially uniform
  and the average conditional density coincides with the unconditional
  density, modulo small-amplitude correlations described by the
  reduced correlation function $\xi(r)$ \citep{book}.  Only in this
  situation may one describe the properties of the force PDF in terms
  of either the $\xi(r)$ function or its Fourier transform, the power
  spectrum \citep{masucci,lattice,book,michael}.  }

\section{The data} 
\label{data} 

We extracted a volume-limited (VL) sample constructed from the data release
7 (DR7) \citep{dr7} of the SDSS (see \citet{sdss_aea,tibor} for
details).  To compute the metric distance, we used the standard
cosmological parameters $\Omega_M=0.3$ and $\Omega_\Lambda=0.7$. We
considered a contiguous sky area with almost uniform coverage and we
applied K-corrections to calculate absolute magnitude $M_r$ in the $r$
band (corrected for Galactic absorption) as in \citet{sdss_aea}.

{ We selected three samples (see Table \ref{tbl}).
\begin{table}
\begin{center}
\begin{tabular}{|c|c|c|c|c|c|c|}
  \hline
  VL sample & $R_{min}$ & $R_{max}$ & $M_{min}$ & $M_{max}$ & $ N$ & $\Lambda$ \\
  \hline
    VL1    & 70  & 450 & -20.8 & -21.8   & 112860 & 1.5 \\
    VL2    & 50  & 200 & -18.9 & -21.1   & 73810  & 0.6 \\
    VL3    & 200 & 600 & -21.6 & -22.8   & 51697  & 2.6 \\
   \hline
\end{tabular}
\end{center}
\caption{{  Main properties of the obtained VL samples with K-corrections
  and without E-corrections: $R_{min}$, $R_{max}$ (in Mpc/h) are the
  chosen limits for the metric distance; ${M_{min}, \,M_{max}}$ define
  the interval for the absolute magnitude in each sample. Finally, $N$
  is the number of galaxies in the sample and $\Lambda$ is the nearest
  neighbor distance (in Mpc/h).}} 
\label{tbl} 
\end{table}
The sample VL1, containing about one fifth of all galaxies in DR7, has
relatively large spatial extensions and small spread in galaxy
luminosity.  On the other hand, the sample VL2 contains fainter
galaxies and covers a smaller volume, while VL3 contains brighter
galaxies and is more extended in radial distance.

In Table \ref{tbl} we also report the average distance between nearest
neighboring galaxies: this provides a lower cut-off for the
computation of real-space statistical properties. It is also directly
related to the large-field tail of the PDF of the force.  The fainter
the galaxies, clearly, the smaller the value of $\Lambda$
\citep{book}.  }

We do not apply any correction for redshift distortion,thus our
results are given in redshift space. We expect the peculiar velocities
to affect the small-scale properties, where their redshift distortions
are relatively important, of the statistical quantities we measured
but to leave conclusions on large scales unchanged.

\section{Methods}
\label{methods} 

The average conditional density in Eq.\ref{apd2} is computed by
evaluating, on the scale $r$, the average of the $N(r)$ density
determinations over the sample points whose minimal distance from
sample boundaries is smaller than $r$.  The force $F$ (Eq.\ref{force})
is computed in a sphere of radius $r$ centered on a galaxy with the
same constraint as above, so that only complete spheres are
considered. This implies that the number of determinations depends on
the scale $r$ \citep{sdss_aea}. We limit the analysis to $N(r) \ge
10^4$. { We note that the force PDF we compute here is intrinsically
  defined for a discrete and stochastic point distribution. As long as
  these points (galaxies) display non-trivial spatial correlations, the
  force PDF deviates from the Holtzmark distribution.}

\section{Results}
\label{results} 

\begin{figure}[htb]
\vspace{1cm}
{
\par\centering \resizebox*{8.0cm}{7cm}{\includegraphics{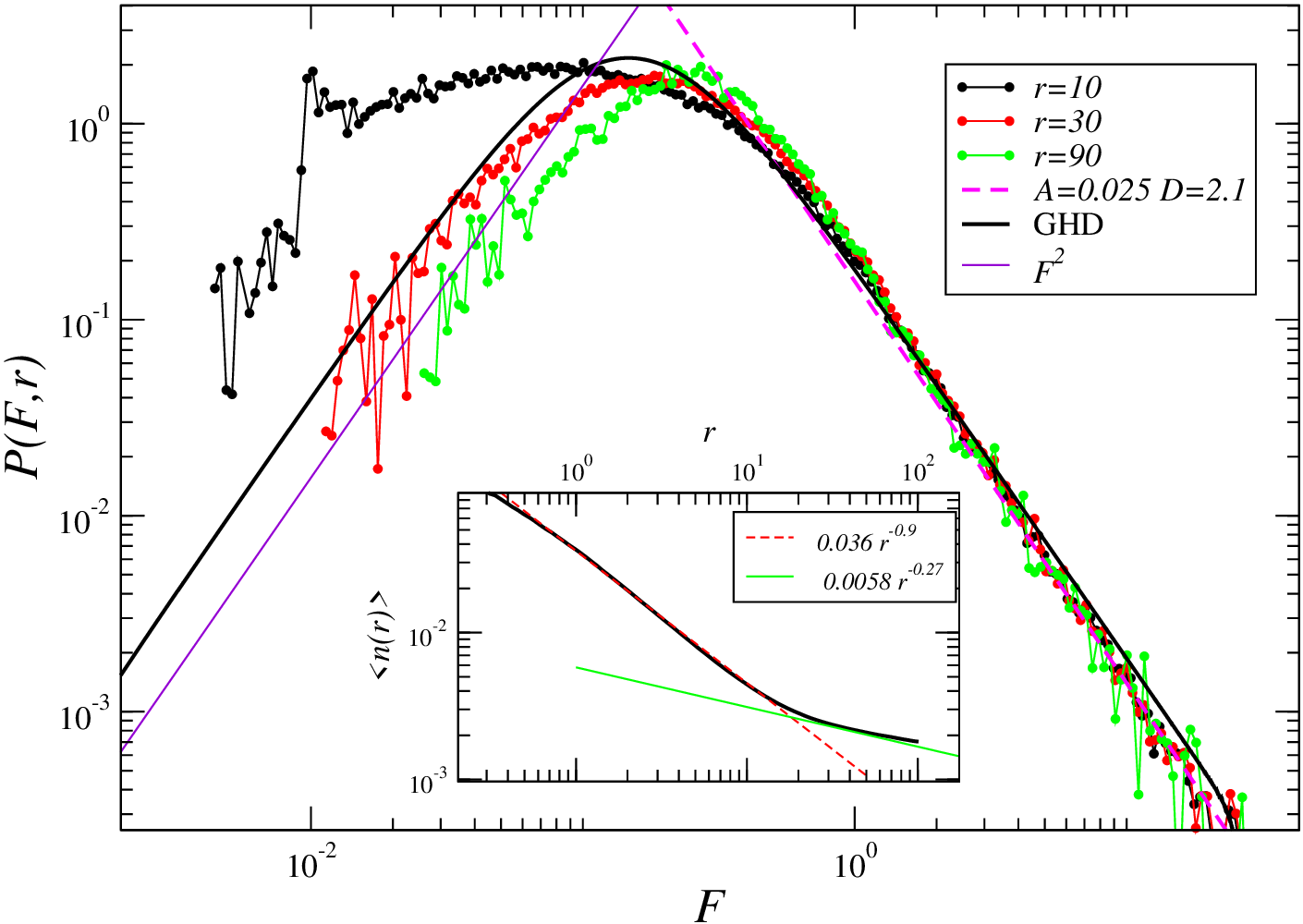}}
\par\centering
}
\caption{PDF of the force for the VL1 computed one different scales
  $r$. The solid line is Eq.\ref{eqHD1} with the parameters from
  Eq.\ref{expfit1}. In the inset panel, the behavior of the average
  conditional density is plotted in addition to its best fits for
  $r<20$ Mpc/h and for $20<r<80$ Mpc/h. \label{fig1}}
\end{figure}

We find that the average conditional density for the VL1 sample (see
Fig.\ref{fig1}, inset panel) behaves for $r<20$ Mpc/h as
\be \langle n(r) \rangle_p = 0.036 \times r^{-0.9} \;,
\label{expfit1} 
\ee 
while for $20<r<80$ Mpc/h we get $\langle n(r) \rangle_p = 0.0058
\times r^{-0.26}$ \citep{tibor}
\footnote{ { The relation between the conditional density and the
    standard two-point correlation function, and in particular the
    relation between their respective exponents in a finite sample is
    discussed in \citet{book} (see pg.240-242)}.}. { This behavior
  implies that a crossover toward homogeneity not being found up to
  $\sim 80 $ Mpc/h. This implies that fluctuations are still large
  when filtered on scales $\sim 80$ Mpc/h and that the standard
  two-point correlation function $\xi(r)$ results, which are biased by
  major systematic effects in these samples, may underestimate the
  amplitude of fluctuations on scales $> 10$ Mpc/h (see the discussion
  in \citet{sdss_aea,2df_aea}).}

\begin{figure}[htb]
\vspace{1cm}
{
\par\centering \resizebox*{8.0cm}{7cm}{\includegraphics{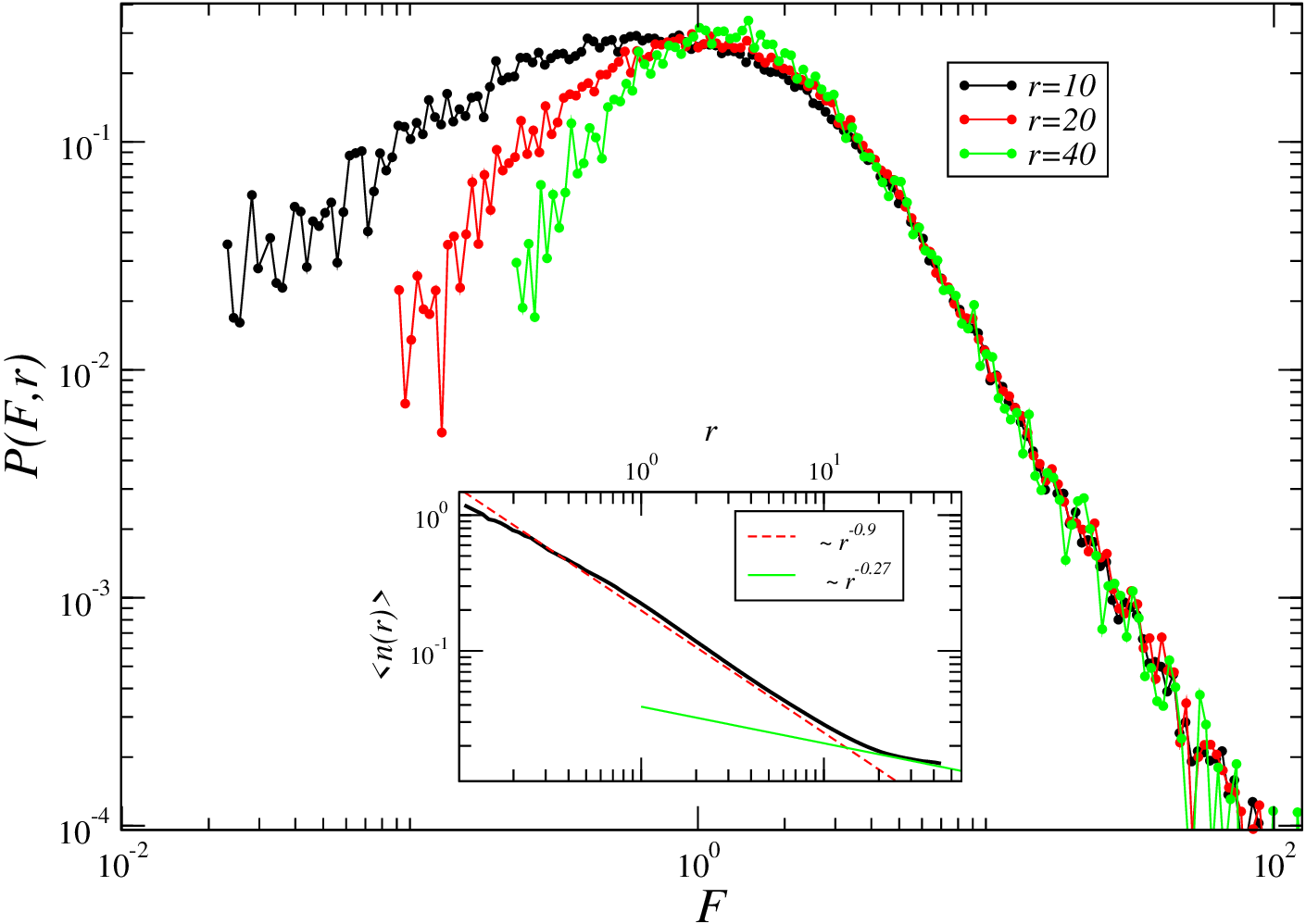}}
\par\centering
}
\caption{The same of Fig.\ref{fig1} but for the sample VL2  \label{fig1b}}
\end{figure}

\begin{figure}[htb]
\vspace{1cm}
{
\par\centering \resizebox*{8.0cm}{7cm}{\includegraphics{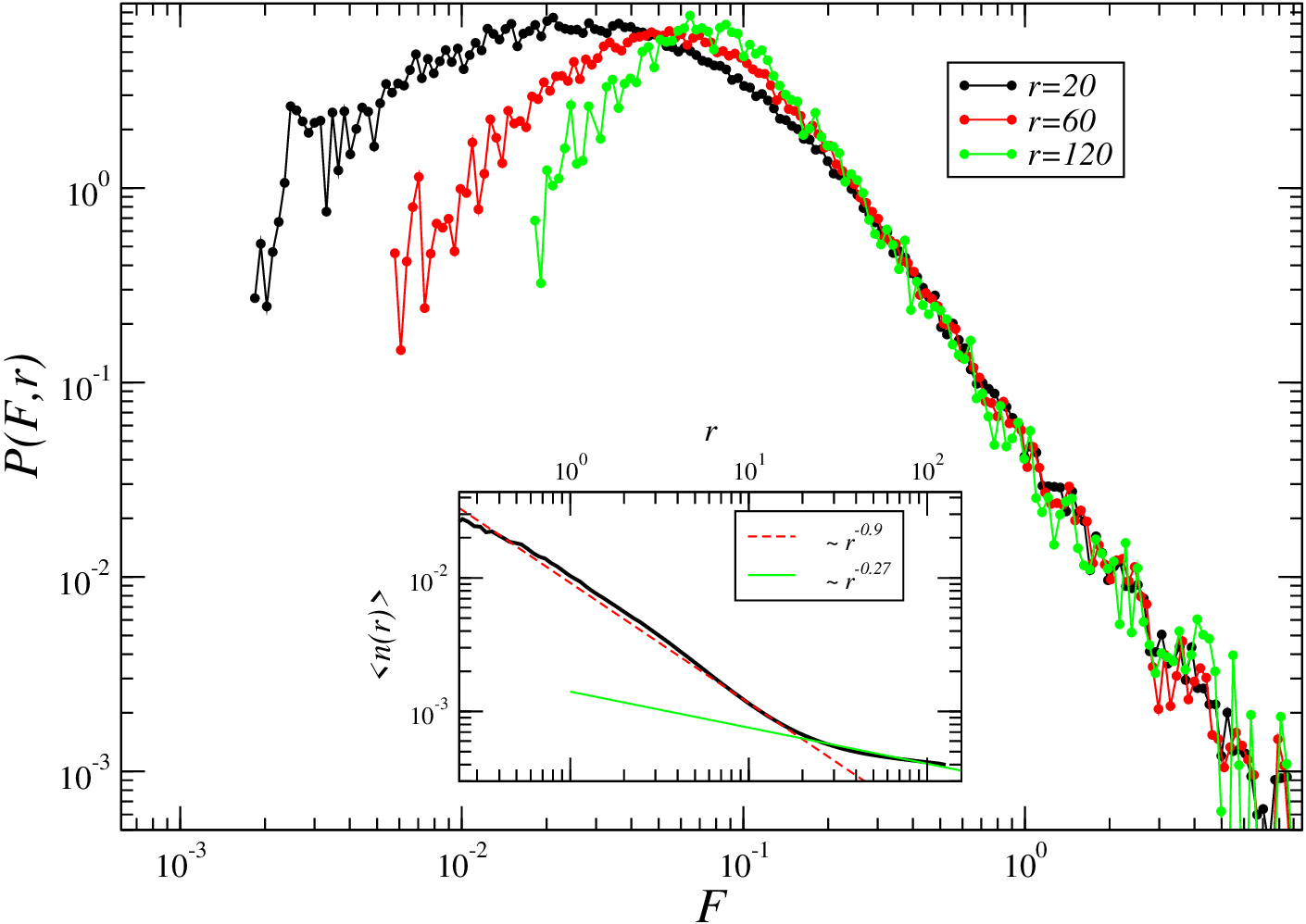}}
\par\centering
}
\caption{The same of Fig.\ref{fig1} but for the sample VL3  \label{fig1c}}
\end{figure}

{ In other two samples considered, VL2 and VL3, we found similar
  behaviors (see Figs.\ref{fig1b}-\ref{fig1c}): (i) the PDF of the
  force does not converge to an asymptotic shape up to $\sim 100$
  Mpc/h and (ii) the conditional density correspondingly does not
  exhibit a well defined trend toward homogenization.  }

In Fig.\ref{fig1}, the PDF of the force for the SDSS-VL1 is shown at
different scales $r$. We note that the large field tail is stable for
all sphere radii considered. It is characterized by a power-law
behavior of the type $ W(F) \sim F^{-\alpha} $ with $\alpha
=-2.1$. From Eq.\ref{eqWnnD1}, we infer that $D=2.2$, thus in
agreement with the value found for the conditional density on small
scales (i.e., Eq.\ref{expfit1}).  Thus the validity of the
approximation at large $F$ values can be simply explained by very
small-scale considerations.
For $r=30$ Mpc/h, the GHD obtained from Eqs.\ref{eqHD1}-\ref{eqHD3}
with the values of $A,D$ measured from the best fit of the conditional
density for $r<20$ Mpc/h, fits the data extremely well.
(i.e. Eq.\ref{expfit1}). In this case, {\it there are no free
  parameters to be adjusted.} In addition, we note that the weak force
tail is proportional to $F^2$, in agreement with Eq.\ref{limits}.

For $r=90$ Mpc/h, the same fit with the GHD, for small $F$ values, is
not as good as for the previous case, while it behaves again as $F^2$
in agreement with Eq.\ref{limits}. This is because the contribution to
the gravitational field from the largest scales in this sample are
non-negligible, and in addition the average conditional density
displays a different slope for $r>20$ Mpc/h than for $r<20$
Mpc/h. This implies that the GHD distribution is only an approximation
to the real PDF. It is indeed clear that the simplifying assumptions
used to derive Eq.\ref{eqHD1}-\ref{eqHD3}, and in particular the
hypothesis that there are no n-point correlations, are too naive with
respect to the complexity of the true galaxy structure.
The value of the field at which the turnover of the PDF occurs,
depends on the sphere radius $r$ determining the cut-off to the force
calculation in the data. This implies that up to the largest sphere
radius considered we do not find a convergence of the force PDF to an
asymptotic behavior. The PDF peak instead shifts toward stronger field
values at larger $r$, implying that greater and greater contributions
are given by the large scale fluctuations.  We note that there is a
factor of three between the PDF peak location at $r=10$ Mpc/h and at
$r=90$ Mpc/h while the peak of the PDF increases by $20\%$ from $r=50$
Mpc/h to $r=90$ Mpc/h. It is surely smaller than from $r=10$ Mpc/h to
$r=50$ Mpc/h but this implies that a non-negligible contribution to
the force field is caused by large-scale fluctuations.

\section{Comparison with mock galaxy catalogs} 
\label{mocks}

\begin{figure}[htb]
\vspace{1cm}
{
\par\centering \resizebox*{8.0cm}{7cm}{\includegraphics{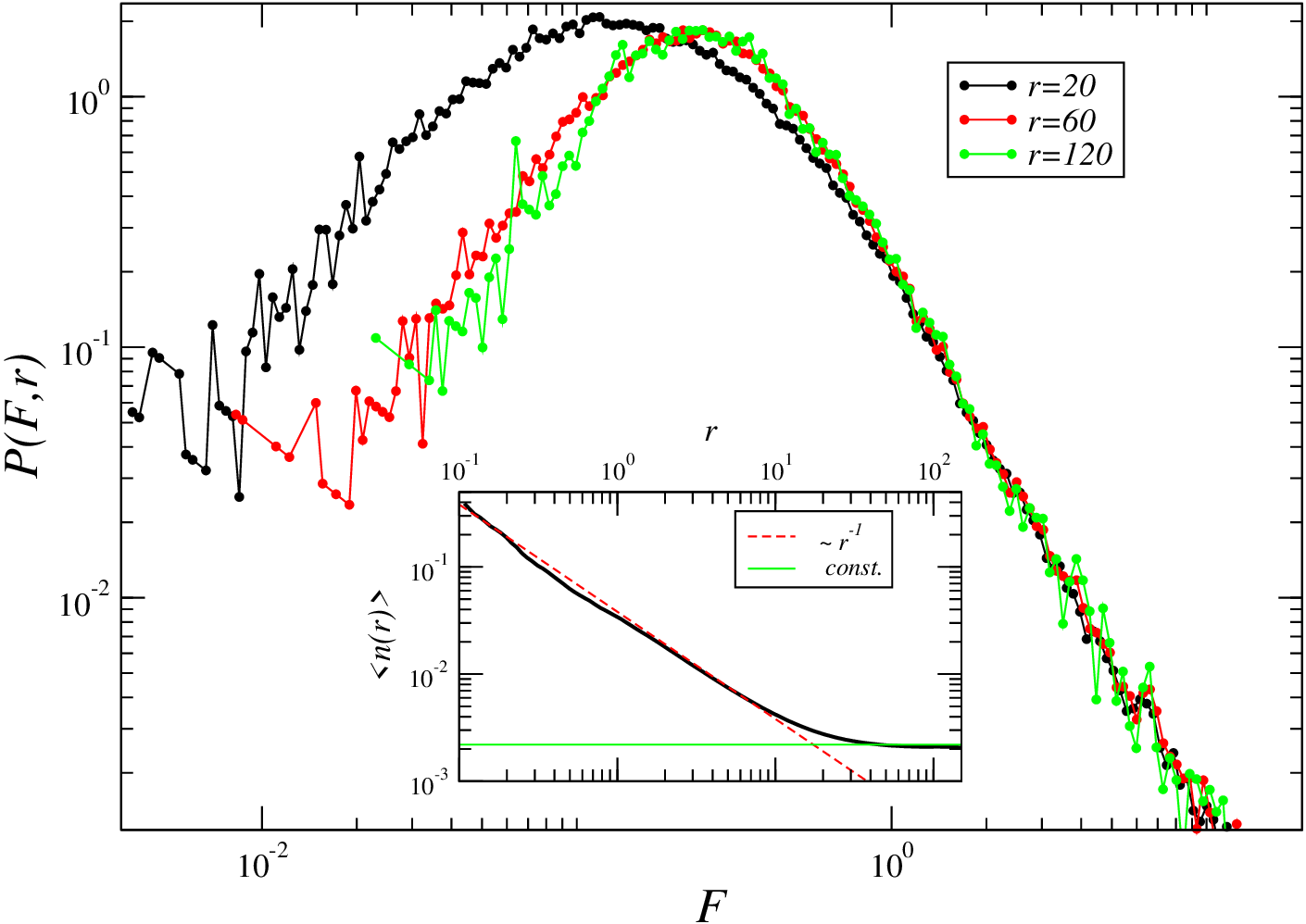}}
\par\centering
}
\caption{{ PDF of the force for the mock sample computed on different
    scales $r$.  In the inset panel, we plot the behavior of the
    average conditional density.}
\label{figmock}}
\end{figure}

{To compare the results obtained in the data with the theoretical
  predictions, we considered a semi-analytic galaxy catalog
  constructed from the Millennium LCDM N-body simulation
  \citep{springel05}. To construct mock samples corresponding to SDSS
  VL samples, we used the full version of the catalog in the $ugriz$
  filter system. The catalog contains about 9 million galaxies in a
  500 Mpc/h cube \citep{croton06} \footnote{See {\tt
      http://www.mpa-garching.mpg.de/galform/agnpaper/} for
    semi-analytic galaxy data files and description, and see {\tt
      http://www.mpa-garching.mpg.de/millennium/} for information on
    Millennium LCDM N-body simulation.}.  We use the absolute
  magnitudes in $r$ filter used in the SDSS case, to construct the
  mock samples with the same limits in absolute magnitude as for the
  SDSS VL samples with K-corrections. As for the real data, we
  considered the redshift-space distribution (more details on the
  construction of mock samples can be found in \citet{sdss_aea}). The
  average distance between the nearest points in the mock galaxy
  sample for which we present results is $\Lambda \sim 1.6 Mpc/h$,
  i.e.  about that of the sample VL1 of SDSS. We note however that in
  other mock samples we found very similar results to the ones
  discussed here.

The main features of the considered statistics in the mock galaxy
samples are (see Fig.\ref{figmock}) (i) beyond a few tens of Mpc the
PDF stabilizes to an asymptotic (sample-independent) shape and (ii)
the conditional density exhibit a well-defined flattening on large
enough scales. Both these behaviors differ from those found in the
real data.  }

{ To conclude this section, we note that in LCDM models, on large
  enough scales, one expects the force PDF to be sample-size
  independent.  In general, if the variance of the force is finite,
  the force PDF is necessarily well defined.  From the definition of
  the force, and after transforming into Fourier space we find that
  \citep{michael}
\[
\langle F^2 \rangle \propto \int d^3k k^{-2} P(k)\;, 
\] 
where $P(k)$ is the power spectrum of the density field.  Therefore, a
sufficient condition for the force to be defined in the usual
thermodynamic limit is just that $k^{-2}P(k)$ be integrable
\citep{michael}. Since we are interested here in only the possible
divergences in this quantity caused by the long distance behavior of
the fluctuations, it is therefore sufficient that we assume that
$\lim_{k \rightarrow 0} kP(k) = 0$. Thus it follows that the force is
well defined in the infinite volume limit if $P(k)$ diverges more
slowly at small $k$ than $k^{-1}$. This condition is satisfied by
standard cosmological density fields for which $P(k) \sim k$ on large
enough scales, i.e.  $r > 100$ Mpc/h \citep{cdm_theo}.  }

\section{Conclusions} 
\label{conclusions}

The gravitational force probability density (PDF) in finite volumes
represents an interesting characterization of the statistical
properties of a given point distribution which is directly related to
{ the statistical properties of density fluctuations and their
  dynamical effects on large scales. By analyzing only the statistical
  properties of the force PDF we have found that this PDF is in
  general related to two-point and higher order correlations.  We have
  measured the PDF in several galaxy samples of the SDSS finding that
  (i) its limiting behaviors can be simply inferred from the galaxy
  correlation properties and that (ii) the PDF does not achieve an
  asymptotic, sample-size-independent, shape within the volumes
  considered, i.e.  $\sim 100$ Mpc/h.  These results agree with the
  studies of the local dipole due to structures, and in particular
  those that do not find a convergence up to $\sim 100$ Mpc/h.

The relation between the force PDF and the dynamics, i.e.  between
gravitational acceleration and peculiar velocity fields requires
several comments. First, we have to assume that galaxies trace the
underlying matter density field, at least on large enough scales
(i.e., that biasing is linear). This is a reasonable assumption
usually employed in the studies of the dipole due to structures
\citep{maller,lavaux2008}.  It is then necessary to assume that
peculiar velocities are parallel to gravitational accelerations.  This
can be verified by linear perturbation theory \citep{pee80} or when
fluctuations are large but the universe is in a curvature-dominated
phase \citep{pwa}. By using these assumptions, we conclude that the
observed values of the bulk flows are {\it compatible} with the
gravitational force fluctuation field from galaxy structures.

In other words,{ these results support the conjecture } that
large-scale structures may be the source of the observed
large-scale flows, which are unexpected in the standard LCDM scenario.
We note that even the correlation properties of the galaxy density field
in the SDSS samples are also at odds with the standard LCDM
predictions \citep{sdss_aea,sdss_epl,2df_epl,2df_aea,sdss_bao}. }

The results of this paper corroborate that the observed large-scale
flows are due to large-scale fluctuations in the galaxy density field.
A more quantitative conclusion must consider the dynamical history of
structure formation and is beyond the scope of this paper.

\acknowledgements I am grateful to Yuri V. Baryshev, Andrea Gabrielli,
Michael Joyce and Nickolay L. Vasilyev for useful discussions and
collaborations. I acknowledge the use of the Sloan Digital Sky Survey
data ({\tt http://www.sdss.org}), of the NYU Value-Added Galaxy
Catalog ({\tt http://ssds.physics.nyu.edu/}) and of the Millennium run
semi-analytic galaxy catalog ({\tt
  http://www.mpa-garching.mpg.de/galform/agnpaper/}).

{}

\end{document}